\documentclass[amsmath,amssymb,prc,superscriptaddress,showpacs,twocolumn]{revtex4}
 
\usepackage{bm}
\usepackage{hyphenat,xspace}
\usepackage{graphicx,epsfig}

\newcommand{\Eq}{Eq.}
\newcommand{\Eqs}{Eqs.}
\newcommand{\Fig}{Fig.}
\newcommand{\Figs}{Figs.}
\newcommand{\Ref}{Ref.}
\newcommand{\Refs}{Refs.}
\newcommand{\Sect}{Sec.}
\newcommand {\mbf}[1]{{\mathbf{#1}}}

\newcommand{\fm}{\;\mathrm{fm}}
\newcommand{\MeV}{\;\mathrm{MeV}}

\newcommand{\zr}{z_R^{-\frac12}}

\newcommand{\ZaR}{\mathcal{Z}_{\alpha R}^{-\frac12}}
\newcommand{\ZbR}{\mathcal{Z}_{\beta R}^{-\frac12}}
\newcommand{\zR}{\mathcal{Z}}

\newcommand{\cm}{\mathrm{c\!\:\!.m\!\:\!.}}
\newcommand{\pd}{p\text{-}d}

\newcommand{\Be}{{}^{10}\mathrm{Be}}
\newcommand{\Bee}{{}^{11}\mathrm{Be}}
\newcommand{\pBe}{p\text{-}{}^{10}\mathrm{Be}}
\newcommand{\pBee}{p\text{-}{}^{11}\mathrm{Be}}

\newcommand{\np}{n\text{-}p}
\newcommand{\pn}{p\text{-}n}

\newcommand{\nBe}{n\text{-}{}^{10}\mathrm{Be}}
\newcommand{\nC}{n\text{-}{}^{12}\mathrm{C}}
\newcommand{\pC}{p\text{-}{}^{12}\mathrm{C}}
\newcommand{\nNi}{n\text{-}{}^{58}\mathrm{Ni}}
\newcommand{\pNi}{p\text{-}{}^{58}\mathrm{Ni}}

\newcommand{\C}{{}^{12}\mathrm{C}}
\newcommand{\Ni}{{}^{58}\mathrm{Ni}}
\newcommand{\dpC}{d\text{+}{}^{12}\mathrm{C}}
\newcommand{\dpNi}{d\text{+}{}^{58}\mathrm{Ni}}
\newcommand{\dpBe}{d\text{+}{}^{10}\mathrm{Be}}

\newcommand{\Beepp}{{}^{11}\mathrm{Be}\text{+}p}

\newcommand{\etal}{{\em et al.}}
 
\begin{document}

\title{Three-body description of direct nuclear reactions: Comparison with 
the continuum discretized coupled channels method}
  
\author{A.~Deltuva} 
\email{deltuva@cii.fc.ul.pt}
\affiliation{Centro de F\'{\i}sica Nuclear da Universidade de Lisboa, 
P-1649-003 Lisboa, Portugal}

\author{A.~M.~Moro} 
\affiliation{Departamento de FAMN, Univ. de Sevilla, Spain}

\author{E.~Cravo} 
\affiliation{Centro de F\'{\i}sica Nuclear da Universidade de Lisboa, 
P-1649-003 Lisboa, Portugal}

\author{F.~M.~Nunes} 
\affiliation{NSCL and Dept. of Physics and Astronomy, Michigan State Univ., 
East Lansing MI 48824, U.S.A.}

\author{A.~C.~Fonseca} 
\affiliation{Centro de F\'{\i}sica Nuclear da Universidade de Lisboa, 
P-1649-003 Lisboa, Portugal}

\received{7 August 2007}

\begin{abstract}
The continuum discretized coupled channels (CDCC) method  is compared to the 
exact solution of the three-body Faddeev equations in momentum space.
We present results for: 
i)~elastic and breakup observables of $\dpC$ at $E_d=56$ MeV, 
ii)~elastic scattering of $\dpNi$ at $E_d=80$ MeV, and 
iii)~elastic, breakup and transfer observables for $\Beepp$ at 
$E_{\Bee}/A=38.4$ MeV. 
Our comparative studies show that, in the first two cases, the CDCC method
is a good approximation to the full three-body Faddeev solution, 
but for the $\Bee$ exotic nucleus, depending on the observable or the kinematic
regime, it may miss out some of the dynamic three-body 
effects that appear through the explicit coupling to the transfer channel. 
\end{abstract}

\pacs{24.10.-i, 24.10.Eq, 25.55.Ci, 25.55.Hp, 25.60.Bx, 25.60Gc, 25.60.Je}

\maketitle


\section{Introduction}\label{Intr}

The strong coupling between elastic and breakup channels in direct nuclear reactions involving 
deuterons led to the development of the continuum discretized coupled channels (CDCC) method 
where an effective three-body problem is solved approximately 
via the expansion of the full wave function in a selected set of continuum wave functions 
of a given pair subsystem Hamiltonian. Initial work by Johnson and Soper \cite{soper} showed that
deuteron breakup was very important to understand reactions involving the deuteron.
In that work, a two channel problem was solved where the deuteron continuum was represented
by a single discrete s-state. Later developments by Rawitscher \cite{rawitscher} and
Austern  \cite{Austern-87} helped to introduce a more realistic representation of the continuum; 
further numerical implementations of the method proved its feasibility \cite{Yahiro-82-86}.
Originally applied to reactions with the deuteron (e.g. \cite{Yahiro-84}),
it has since been extended to describe reactions with radioactive nuclear beams
(e.g. \cite{Tostevin-01,Tostevin-02,Shrivastava-04,Moro-03,Jeppesen-06}),
namely to study elastic, transfer and breakup cross sections that result from the collision 
of a halo nucleus with a proton or a stable heavier target such as $\C$ or ${}^{208}\mathrm{Pb}$.  

In a recent paper \cite{Moro-06} CDCC results obtained with two different basis sets, 
namely the basis set using the continuum of the projectile in the entrance channel, 
and the one using the continuum of the composite system in the final transfer channel, 
led to  substantially different breakup cross sections for $p(\Bee, \Be)pn$. 
These findings raise concern about the accuracy of the CDCC method, not only as a means to describe 
reaction dynamics, but also as an accurate tool to extract structure information on halo nuclei.

An alternative approach to the solution of effective three-body problems is the solution 
of the Faddeev equations \cite{Faddeev-60-61} for the wave function components or the equivalent 
Alt, Grassberger and Sandhas (AGS) equations \cite{Alt-67} for the transition operators. 
The application of exact Faddeev/AGS equations to the study of direct nuclear reactions 
has been shadowed in the past by the difficulty in dealing with the long range Coulomb force 
between charged particles \cite{Alt-02}. 
In recent calculations for the reaction $\dpC$ \cite{Alt-07} separable potentials are used 
and the Coulomb is taken into account only approximately. 
Given the progress achieved recently \cite{Deltuva-05-05-05} for $\pd$ elastic scattering and breakup 
we can now address the solution of effective three-body systems where two of the particles have charge.
This was done first for $d\text{-}\alpha$ elastic scattering and breakup \cite{Deltuva-06} 
and later for $\pBee$ elastic scattering and breakup \cite{Deltuva-06a} where $\Bee$ is a halo nucleus 
made up of a neutron and an inert $\Be$ core. More recently the same system was used to study the 
convergence of the Faddeev/AGS multiple scattering series \cite{crespo:accept} at intermediate 
energy as a means to test the Glauber method.
In all these works \cite{Deltuva-05-05-05,Deltuva-06,Deltuva-06a,crespo:accept} we solve the
AGS equations without resorting to a separable representation of the underlying interactions.
Therefore the corresponding two-vector-variable integral equations are numerically solved without
any approximations beyond the usual partial-wave decomposition and discretization of momentum meshes.

Although CDCC was initially  introduced  
as a practical way of solving a complicated three-body  
scattering problem through a set of coupled  
Schr\"odinger-like equations, later works \cite{Austern-89,Austern-96} 
tried to obtain a more formal  justification of the method, by relating it 
to a truncation of an orderly set of Faddeev equations.  Furthermore, it is 
argued that the  CDCC 
solution approaches the exact (Faddeev) solution as the model space is 
increased. 
Although qualitative arguments are provided in those works to support the conclusions, they
lack a numerical comparison between the CDCC and Faddeev methods in 
specific cases. The possibility of performing this comparison, thanks to 
the recent developments in the numerical implementation of the AGS 
equations, is another motivation for the present work. 


Given these important new developments we propose here to benchmark, in a few test cases, CDCC results 
with exact solutions of the AGS equations. For this comparison, we have selected 
the reactions  $\dpC$ , $\dpNi$ and $\Beepp$. 
The first two reactions correspond to classic cases for which there is data available
\cite{Matsuoka-86,Duhamel-71,Stephenson-83}. 
The third reaction involves the scattering of a halo nucleus on a very light target, for which 
elastic, breakup and transfer have been measured before \cite{Lapouxsub,Shrivastava-04,Fortier-99}.

In Section~\ref{sec:AGSformalism} we present the AGS formalism, and 
in Section~\ref{sec:differentCDCC} we outline the different CDCC methodologies we use. 
In Section~\ref{sec:calculations} we describe the details of the calculations and 
in Section~\ref{sec:results} the results are presented. 
Conclusions are given in Section~\ref{sec:conclusions}.  


\section{Three-body equations} \label{sec:AGSformalism}  

This section provides the theoretical framework on which we base our calculations. 
Our treatment of the Coulomb interaction \cite{Deltuva-05-05-05} relies on  
the screening and renormalization techniques proposed in \Ref~\cite{Taylor-74-75} for two charged 
particle scattering and extended in \Ref~\cite{Alt-78-80} to three-particle scattering. 
The Coulomb potential is screened, standard scattering theory for short-range potentials is used, 
and the renormalization procedure is applied to obtain the results for the unscreened limit.

\begin{figure}[th!b]
\begin{center}
\includegraphics[scale=0.9]{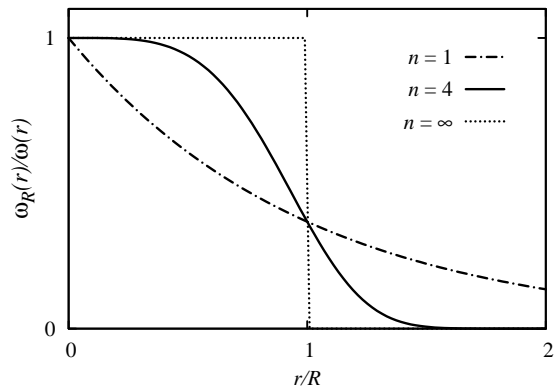}
\end{center}
\caption{\label{fig:screenf} Screening function $w_R(r)/w(r)$ as function of the distance 
between the two charged particles $r$ for characteristic values of the parameter $n$ in \Eq~\eqref{eq:wr}: 
$n=1$ (dashed-dotted curve) corresponds to Yukawa screening,
$n=4$ (solid curve) is the choice of this 
paper, and $n \to \infty$ (dotted curve) corresponds to a sharp cutoff. }
\end{figure}

In the traditional odd-man-out notation of the three-body problem where pair $(\beta, \gamma)$ is 
denoted by $\alpha$  $\, (\alpha, \beta, \gamma \equiv 1,2,3)$, the Coulomb potential $w_{\alpha R}$ 
is screened around the separation $r=R$ between two charged baryons $\beta$ and $\gamma$.
We choose $w_{\alpha R}$ in configuration space as 
\begin{equation} \label{eq:wr}
w_{\alpha R}(r) = w_{\alpha}(r) \; e^{-(r/R)^n},
\end{equation}
where $w_{\alpha}(r) = \alpha_e\, Z_{\beta}\, Z_{\gamma}/r$ represents the true Coulomb potential, 
with $Z_{\beta}$ ($Z_{\gamma}$) being the atomic number of particle $\beta$ ($\gamma$), 
$\alpha_e \approx 1/137$ the fine structure constant, and $n$ controlling the smoothness of the screening. 
We prefer to work with a sharper screening than the Yukawa screening $(n=1)$ of \Ref~\cite{Alt-02}. 
We want to ensure that the screened Coulomb potential $w_{\alpha R}$ approximates well 
the true Coulomb potential $w_{\alpha}$ for distances $r<R$ and simultaneously vanishes rapidly 
for $r>R$, providing a comparatively fast convergence of the partial-wave expansion. 
The screening functions for different $n$ values are compared in \Fig~\ref{fig:screenf}, 
showing that the choice $n=4$ includes much more of the exact Coulomb potential at short distances 
than the Yukawa screening. 
In contrast, the sharp cutoff $(n \to \infty)$ yields an unpleasant oscillatory behavior 
in the momentum-space representation, leading to convergence problems. 
In \Ref~\cite{Deltuva-05-05-05} we found the values $3 \le n \le 6$ to provide a sufficiently smooth,
but at the same time a sufficiently rapid screening around $r=R$;
$n=4$ is our choice in the present paper.

We solve the AGS three-particle scattering equations~\cite{Alt-67} in momentum space
\begin{subequations}\label{eq:AGS}
  \begin{align} \label{eq:Uba}
     U^{(R)}_{\beta \alpha}(Z) = {} & \bar{\delta}_{\beta \alpha} G_0^{-1}(Z)
     + \sum_{\sigma} \bar{\delta}_{\beta \sigma} T^{(R)}_\sigma (Z) G_0(Z)
     U^{(R)}_{\sigma \alpha}(Z), \\ \label{eq:U0a}
     U^{(R)}_{0 \alpha}(Z) = {} & G_0^{-1}(Z)
     + \sum_{\sigma}  T^{(R)}_\sigma (Z) G_0(Z) U^{(R)}_{\sigma \alpha}(Z),
  \end{align}
\end{subequations}
where $\bar{\delta}_{\beta \alpha} = 1 - {\delta}_{\beta \alpha}$,
$ G_0(Z)$ is the free resolvent, and 
$T^{(R)}_\sigma (Z)$ the two-particle transition matrix
derived from nuclear plus screened Coulomb potentials
\begin{equation}\label{eq:TR}
T^{(R)}_\alpha (Z) = (v_\alpha + w_{\alpha R}) + 
     (v_\alpha + w_{\alpha R}) G_0(Z) T^{(R)}_\alpha (Z),
\end{equation}
embedded in three-body space. 
The operators $U^{(R)}_{\beta \alpha}(Z)$ and $U^{(R)}_{0 \alpha}(Z)$ 
are the three-particle transition operators for elastic/rearrangement and breakup scattering respectively; 
their dependence on the screening radius $R$ is notationally indicated.
On-shell matrix elements of the operators \eqref{eq:AGS} between two- and three-body channel states 
$|\phi_\alpha (\mbf{q}_i) \nu_{\alpha_i} \rangle$ and 
$|\phi_0 (\mbf{p}_f \mbf{q}_f) \nu_{0_f} \rangle$ with discrete quantum numbers $\nu_{\alpha_i}$,
Jacobi momenta $\mbf{p}_i $ and $\mbf{q}_i $, energy $E_{\alpha i}$, and $Z=E_{\alpha i}+i0$, 
do not have a $R \to \infty$ limit.
However, as demonstrated in \Refs~\cite{Deltuva-05-05-05,Alt-78-80}, the three-particle amplitudes 
can be decomposed into long-range and Coulomb-distorted short-range parts, 
where the quantities diverging in that limit are of two-body nature, i.e., 
the on-shell transition matrix
\begin{equation} \label{eq:Tcm}
T^{\cm}_{\alpha R}(Z) = W^{\cm}_{\alpha R} + 
W^{\cm}_{\alpha R} G^{(R)}_{\alpha}(Z) T^{\cm}_{\alpha R}(Z),
\end{equation}
\begin{equation} \label{eq:GRa}
 G^{(R)}_\alpha (Z) = (Z - H_0 - v_\alpha - w_{\alpha R})^{-1},
\end{equation}
derived from the screened Coulomb potential between spectator and the center of mass (c.m.) of the bound 
pair, the corresponding wave function, and the screened Coulomb wave function for the relative motion 
of two charged particles in the final breakup state.
Those quantities, renormalized according to \Refs~\cite{Deltuva-05-05-05,Alt-78-80}, in the 
$R \to \infty$ limit converge to the two-body Coulomb scattering amplitude 
$\langle\phi_\alpha (\mbf{q}_f)\nu_{\alpha_f}|T^{\cm}_{\alpha C} 
|\phi_\alpha (\mbf{q}_i)\nu_{\alpha_i}\rangle$ (in general, as a distribution) 
and to the corresponding Coulomb wave functions, respectively,
thereby yielding the three-particle scattering amplitudes in the proper Coulomb limit
\begin{subequations}\label{eq:UC}
\begin{gather} \label{eq:UC1}
  \begin{split}
    \langle  \phi_\beta (\mbf{q}_f)  \nu_{\beta_f} & | U_{\beta \alpha}
    |\phi_\alpha (\mbf{q}_i) \nu_{\alpha_i} \rangle   \\ = {}&
    \delta_{\beta \alpha}
    \langle \phi_\alpha (\mbf{q}_f) \nu_{\alpha_f} |T^{\cm}_{\alpha C}
    |\phi_\alpha (\mbf{q}_i) \nu_{\alpha_i} \rangle  \\ & +
    \lim_{R \to \infty} \{ \ZbR(q_f)
    \langle \phi_\beta (\mbf{q}_f) \nu_{\beta_f} |
            [ U^{(R)}_{\beta \alpha}(E_{\alpha i} + i0)  \\ & -
              \delta_{\beta\alpha} T^{\cm}_{\alpha R}(E_{\alpha i} + i0)]
            |\phi_\alpha (\mbf{q}_i) \nu_{\alpha_i} \rangle
            \ZaR(q_i) \},
  \end{split} \\  \label{eq:UC0}
    \begin{split}
      \langle \phi_0  (\mbf{p}_f \mbf{q}_f) & \nu_{0_f}  | U_{0 \alpha}
      |\phi_\alpha (\mbf{q}_i) \nu_{\alpha_i} \rangle  \\ = {}&
      \lim_{R \to \infty} \{ \zr(p_f)
      \langle \phi_0 (\mbf{p}_f \mbf{q}_f) \nu_{0_f} | \\ & \times
      U^{(R)}_{0 \alpha}(E_{\alpha i} + i0)
      |\phi_\alpha (\mbf{q}_i) \nu_{\alpha_i} \rangle \ZaR(q_i) \}.
    \end{split}
  \end{gather}
\end{subequations}
The renormalization factors $\zR_{\alpha R}(q_i)$ and $z_R(p_f)$ are diverging phase factors 
given in \Refs~\cite{Deltuva-05-05-05,Taylor-74-75,Alt-78-80}.
\begin{subequations} \label{eq:zrq}
  \begin{gather}
    \zR_{\alpha R}(q) = e^{-2i \delta_{\alpha R}(q)},
  \end{gather}
  where $\delta_{\alpha R}(q)$, though independent of the relative  
  angular momentum $l$ in the infinite $R$ limit, is realized by
  \begin{gather}    \label{eq:phiRl}
    \delta_{\alpha R}(q) = \sigma^{\alpha}_l(q) -\eta^{\alpha}_{lR}(q), 
  \end{gather}
with the diverging screened Coulomb phase shift $\eta^{\alpha}_{lR}(q)$ corresponding to 
standard boundary conditions and the proper Coulomb one $\sigma^{\alpha}_l(q)$ 
referring to the logarithmically distorted Coulomb boundary conditions in channel $\alpha$ 
with orbital angular momentum $l$ between particle-pair $\alpha$. For the screened Coulomb 
potential of \Eq~\eqref{eq:wr} the infinite $R$ limit of $\delta_{\alpha R}(q)$ 
is known analytically 
\begin{gather}  \label{eq:phiRlln}
  \delta_{\alpha R}(q) =  \kappa_{\alpha}(q)[\ln{(2qR)} - C/n],
  \end{gather}
\end{subequations}
$\kappa_{\alpha}(q) = \alpha_e \,  Z_{\alpha}(Z_{\beta} + Z_{\gamma}) M_{\alpha}/q$ 
being the Coulomb parameter, $M_{\alpha}$ the reduced mass, and 
$C \approx 0.5772156649$ the Euler number.
Likewise
 \begin{subequations} \label{eq:zrp}
  \begin{gather}
    z_R(p) = e^{-2i \delta_{R}(p)},
  \end{gather}
  where
  \begin{gather}    \label{eq:phiRp}
    \delta_{R}(p) = \kappa(p)[\ln{(2pR)} - C/n] 
  \end{gather}
\end{subequations}
with $\kappa(p) = \alpha_e \,  Z_{\beta} \, Z_{\gamma} \mu_{\alpha}/p$ where $\beta$ and $\gamma$ 
denote the two charged particles and $\mu_{\alpha}$ their respective reduced mass.
The $R \to \infty$ limit in \Eqs~\eqref{eq:UC} has to be calculated numerically, 
but due to the short-range nature of the corresponding operators it is reached 
with sufficient accuracy at rather modest $R$ if the form of the screened Coulomb potential
has been chosen successfully as discussed above.
More details on the practical implementation of the screening and renormalization approach 
are given in \Ref~\cite{Deltuva-05-05-05}.

The three-body results are obtained from the solution of the AGS equations \eqref{eq:AGS} for 
nuclear plus screened Coulomb interaction together with the renormalization procedure \eqref{eq:UC}. 
The equations are solved using partial wave decomposition and retaining as many channels 
as needed for convergence. Our numerical technique for solving AGS equations with non-separable
potentials is explained in more detail in \Refs~\cite{deltuva:phd,deltuva:03a} in the context of
nucleon-deuteron scattering.


\section{CDCC Formalism} \label{sec:differentCDCC}  

The CDCC method \cite{Austern-87,Yahiro-82-86} was introduced as an approximate solution to the 
three-particle Schr\"odinger equation. Its main objective 
is to provide a reliable yet practical way of describing reactions involving three-body breakup. 

Let us consider specifically the breakup reaction $p+t \rightarrow c+x+t$.
In CDCC, the wavefunction is expanded in terms of only one Jacobi coordinate set $(\bm r, \bm R)$, 
\begin{equation} \label{cdccwf0} 
\Psi_{\bm K}({\bm r},{\bm R}) = 
\sum_{p} \phi_{p}({\bm r}) \psi^{\bm K}_{p}({\bm R}) + 
\int_0^\infty {d \bm k} \phi_{\bm k}({\bm r}) \psi^{\bm K}_{\bm k}({\bm R}), 
\end{equation} 
where $\phi_{p}({\bm r})$ are eigenfunctions of the projectile, $p({\bm k})$ being a general subscript 
for projectile bound (continuum) states, and $\psi^{\bm K}_{p}({\bm R})$ the spectator wavefunction 
for the motion of the projectile relative to the target. 
The Jacobi coordinate $\bm r$ describes the $c+x$ relative motion while $\bm R$ 
the $p+t$ relative motion. 
In CDCC the Schr\"odinger equation 
expressed in this Jacobi set reads
\begin{equation} \label{schro3b} 
(H_{3b}-E) \Psi_{\bm K}({\bm r},{\bm R})=0. 
\end{equation}
where the three-body Hamiltonian is separated into the internal Hamiltonian of the projectile and
the relative motion between the projectile and the target:
$H_{3b} = H_{int} + T_R + U_{xt} + U_{ct}$, where $H_{int} = T_r + V_{xc}({\bm r})$. 
The projectile is modelled by a real potential which produces its initial bound state, 
whereas the fragment-target interactions should contain absorption from channels 
not included explicitly in the model (optical potentials). 
In principle, this is important for the validity of the CDCC method \cite{Austern-89-96}, 
since it reduces the coupling to the other three-body channels best described by other Jacobi 
sets taken explicitly into account by the Faddeev method. 

For practical reasons, the integral over projectile scattering states \eqref{cdccwf0}
is discretized and truncated at a maximum energy. 
There are several methods of discretization but here we use the average method, 
where the $c+x$ scattering radial functions $u_{ k}(r)$ are averaged over $k$ 
to be made square integrable \cite{Tostevin-01,Moro-06}. 
Thus, the radial functions for the continuum bins in the average method, $\tilde u_{p}(r)$, 
are a superposition of the projectile scattering eigenstates 
\begin{equation} \label{bins}
\tilde u_{p}(r) = \sqrt{\frac{2}{\pi N_p }} \int_{k_{p-1}}^{k_{p}} g_p(k) u_{ k}(r)\, dk ,
\end{equation}
with weight function $g_p(k)$. 
The normalisation constant is defined by $N_p = \int_{k_{p-1}}^{k_{p}} |g_p (k)|^2 \, dk$.

After a few steps of algebra and using the eigenvalue equation for the projectile 
$H_{int} \phi_p = \varepsilon_p \phi_p$, the standard CDCC equation becomes 
\begin{equation} \label{cdcc-simple2} 
[T_R + V_{pp}(R) - E_{p}] \psi^{\bm K}_{p}({\bm R}) = 
- \sum_{p' \ne p} V_{pp'}(R) \psi^{\bm K}_{p'}({\bm R}),
\end{equation}
where $E_{p}= E_{\cm} - \varepsilon_{p}$, 
and the coupling potentials contain both nuclear and Coulomb parts 
$V_{pp'}(R)=\langle \phi_p | U_{xt} + U_{ct} | \phi_{p'} \rangle$. 
This equation is indeed a coupled channel equation, coupling the projectile ground state 
to its continuum states via $V_{0 p},$ but also coupling projectile states within the continuum, 
called the continuum-continuum couplings. 
The solution of the coupled equations provide the wavefunctions 
$\psi^{\bm K}_{\bm k}({\bm R})$. The 
scattering observables (associated to the elastic and breakup channels) are 
extracted from the asymptotic behaviour of $\psi^{\bm K}_{\bm k}({\bm R})$.
Numerical solutions of Eq. (\ref{cdcc-simple2})
involve partial wave expansions of $\phi_{p}(\bm r)$, $\psi^{\bm K}_{\bm k}({\bm R})$ and a multipole
decomposition of $U_{xt} + U_{ct}$.

In its standard form, CDCC models the breakup of a projectile as inelastic excitation (CDCC-BU).
However, the CDCC wavefunction of \Eq~\eqref{cdccwf0} can also be used in the exit channel 
of a transfer reaction. 
In that case, the breakup process is understood as a transfer of the fragment to the continuum 
of the final composite system (CDCC-TR*). 
In the CDCC-TR* scheme, the scattering observables can be obtained inserting 
the CDCC wavefunction in the prior form of the transition amplitude:
\begin{equation} 
T_{prior}=
\langle \Psi _{f}^{(-)}| U_{xt}+U_{ct}-U_{pt} | \phi_{p} \chi_{p} \rangle ,
\label{Tprior}
\end{equation}
where  $\chi_{p}$ is a distorted wave, generated by the potential $U_{pt}({\bm R})$. 
In this work, we chose $U_{pt}({\bm R}) = \langle \phi_0 | U_{xt}+U_{ct} |\phi_0 \rangle $ 
as the single folding of the core-target and the fragment-target interactions
over the projectile's ground state (the Watanabe potential).
We notice that, if $\Psi _{f}^{(-)}$ is the exact 
solution of the three-body Hamiltonian,  the transition amplitude (\ref{Tprior})
is exact and  does not depend on the choice of the auxiliary potential $U_{pt}$.
In  practice, this wavefunction is replaced by an approximate one which, 
in the TR* approach, corresponds to the CDCC expansion in the final channel.

The transfer to the continuum approach seems to provide a good description 
of the data in some cases \cite{Moro-03a,Escrig-07,Jeppesen-06}. 
In Ref.~\cite{Moro-06}, a detailed discussion of these two mechanisms is 
presented  along with a comparative study.
In principle,  and as long as the CDCC model space is sufficiently large, one would expect that
different choices of the Jacobi coordinate produce the same results. 
However, in \cite{Moro-06} it is shown that this equivalence does not hold for a number of 
reaction observables.


\section{Details of the calculations} \label{sec:calculations}

As mentioned above we study the scattering of deuterons on $^{12}\mathrm{C}$ at $E_d=56\MeV$
and $^{58}\mathrm{Ni}$ at $E_d=80\MeV,$ as well as $^{11}\mathrm{Be}$ on protons 
at $E_{\text{Lab}}/A = 38.4\MeV.$ 
All reactions are considered as effective three-body problems, namely $p + n+ ^{12}\mathrm{C}$, 
$p + n+ ^{58}\mathrm{Ni}$, and $^{10}\mathrm{Be} + n + p$, where $^{12}\mathrm{C}$, 
$^{58}\mathrm{Ni}$ and $^{10}\mathrm{Be}$ are taken as inert cores. 
Therefore in the present section we define the interactions between all pairs 
together with the model space used in solving the AGS and the CDCC equations.
For simplicity all interactions are taken spin-independent
as often done in many CDCC calculations (e.g. \Ref~\cite{Austern-87})
and therefore all particles are considered spinless bosons.
Nevertheless, spin-independent interactions, especially for the $\np$ system,
are only semirealistic, but are sufficient for a benchmark comparison. This
is not a limitation of the CDCC method nor of the AGS method as
we have already demonstrated in nuclear reaction calculations
\cite{Deltuva-06,Deltuva-06a,crespo:accept} with realistic spin-dependent potentials.

\subsection{$\bm\dpC$ at $\bm{E_d = 56\MeV}$} \label{subsec:d-c} 

The interactions between neutron-$^{12}\mathrm{C}$ and proton-$^{12}\mathrm{C}$ are
optical potentials that fit the elastic scattering at half the incident laboratory energy 
\cite{johnson72}; the parameters are taken from the global fit by Watson {\it et al.} \cite{Watson-69}.
The neutron-proton bound and continuum states are modelled with a simple Gaussian interaction 
fitted to the deuteron binding energy
\begin{equation} \label{eq:1V(r)}
V(r) = - V_0 \; e^{-(r/r_0)^2} ,
\end{equation}
where $V_0 = 72.15\MeV$ and $r_0 = 1.484\fm.$
The same interaction is used in all three test cases and corresponds to the choice 
in \cite{Matsuoka-82,Yahiro-84}.

The model space needed for converged solutions of the AGS equations contains partial waves 
$l \le 3$ in the $\np$ relative motion, $l \le 12$ in the $\nC$ channel, and 
$l \le 24$ for elastic scattering ($l \le 32$ for breakup) in the $\pC$ channel. 
This last channel is more demanding due to the presence of the Coulomb force. 
Total angular momentum up to $J = 30$ ($J = 60$ for breakup) is included.
The Coulomb potential is screened with a radius of $R=10\fm$ ($R=18\fm$ for breakup) 
and smoothness $n=4$ (see \Eq~\eqref{eq:wr}). 
The exception is the breakup kinematical situations characterized by small momentum transfer in
the $\pC$ subsystem which are sensitive to the Coulomb interaction at larger distances
and therefore need larger screening radius and a special treatment as described in the Appendix.
Note that, as optical potentials are used for $\nC$ and $\pC$, there is no transfer to
$p\text{-}{}^{13}\mathrm{C}$ and $n\text{-}{}^{13}\mathrm{N}$ channels.

The corresponding CDCC calculations include $\np$ partial waves $l \le 8$ 
and bins up to $E_{\text{max}}=46\MeV$ (with 15 (10) energy bins for the even (odd) partial 
waves, 
evenly spaced in linear momentum) integrated up to $R_{\text{bin}}= 80\fm$; 
for the total angular momentum we include $J \le 60$. The
coupling potentials were expanded in multipoles ($Q$) up to  $Q_\mathrm{max} =6$. 
We note that this relatively  
large model space is required to achieve sufficient accuracy for the breakup observables. If 
only elastic scattering is required, $l \le 2$ gives almost a converged result.  
The CDCC equations are solved up to $R_{\text{max}}= 100\fm$.

\subsection{$\bm\dpNi$ at $\bm{E_d = 80\MeV}$} \label{subsec:d-ni} 

Similarly to the Carbon test case, we study the scattering of deuterons from a Ni target
using neutron-$^{58}\mathrm{Ni}$ and proton-$^{58}\mathrm{Ni}$ optical potentials 
from the global parameterization of Becchetti and Grenless \cite{Becchetti-69},
evaluated at half the incident laboratory energy. The $\np$ interaction is  
given by \Eq~\eqref{eq:1V(r)}.

In this case, the model space needed for converged solutions of the AGS equations 
for elastic scattering contains partial waves $l \le 3$ in the $\np$ relative motion, 
$l \le 14$ in the $\nNi$ channel, and $l \le 32$ in the $\pNi$ channel. 
Again this last channel is more demanding due to the presence of the Coulomb force. 
Total angular momentum up to $J = 60$ is included.
The Coulomb potential is screened with a radius of $R=10\fm$ and $n=4$ (see \Eq~\eqref{eq:wr}).

For the CDCC calculations we include $\np$ partial waves $l \le 2$. For $l=0,1$ the 
continuum was truncated at $E_{\text{max}}=30\MeV$, and divided into 12 bins evenly 
spaced in the linear momentum, whereas for $l=2$ we include excitation energies up to 
$E_{\text{max}}=50\MeV$, and use 20 bins; $Q \leq 2$ multipoles are retained in the expansion 
of the coupling potentials. The coupled equations are integrated up to  
$R_{\text{max}}= 80 \fm$ with total angular momentum up to $J=100$.

\subsection{$\bm\Beepp$} \label{subsec:pBe} 

To describe the scattering of $\Bee$ from a proton target, we need a binding potential 
for the $\nBe$ pair, as well as fragment-target optical potentials. 
The $\nBe$ interaction takes the standard Woods-Saxon form
\begin{equation} \label{eq:2V(r)}
V(r) = - V_0 \; f(r, R_0, a_0) 
\end{equation}
with
\begin{equation} \label{eq:f(r,R,a)}
f(r,R,a) = {\left(1 + e^{(r-R)/a}\right)}^{-1} ,
\end{equation}
where $R_i = r_i A^{\frac{1}{3}}$ and $A$ the mass number of $\Be.$ 
The geometry of the interaction is fixed, with a radius $r_0=1.39\fm$ and a diffuseness $a_0=0.52\fm.$ 
The depth of the interaction is $L$-dependent and the corresponding values of $V_0$ 
for each partial wave are given in Table \ref{Tab:n-10Be}, together with the energies 
for the corresponding bound states and resonance. 
In the three-body calculation the lowest (Pauli forbiden) bound state $|b_0 \rangle $ in $L=0$ 
is moved to a large positive energy $\Gamma$, replacing the potential  $V$ by
$V' = V +|b_0 \rangle \Gamma  \langle b_0|$. In the  $\Gamma \to \infty$ limit this is
equivalent to projecting $|b_0 \rangle $ out as demonstrated in \Ref~\cite{schellingerhout:93a}.
In practical calculations we found that $\Gamma \approx 2$ GeV is sufficiently large
in order to obtain $\Gamma$-independent results.
Thus, the state with $\epsilon^*_0 = -0.503\MeV$ is left as the ground state of $\Bee$. 
The unphysical deep $s$-state is also left out of the CDCC calculations.
\begin{table}[th!b]
\caption{\label{Tab:n-10Be} Parameters of the $\nBe$ interaction in different partial waves 
and resulting energies of bound states and resonance.}
\begin{ruledtabular}
\begin{tabular}{cccc}
$L$  & $V_0$ (MeV) & $\epsilon_L$ (MeV)   & $\epsilon^*_L$ (MeV) \\ \hline
$0$  & 51.639      & $-30.28$             & $-0.503$             \\ 
$1$  & 26.264      & $-0.183$             &                      \\ 
$2$  & 51.639      & $1.317 - i\,0.188/2$ &                      \\ 
$>3$ & 51.639      &                      &                      \\ 
\end{tabular}
\end{ruledtabular}
\end{table}

As for the projectile-target interactions, given such a light proton target, one of the potentials is 
simply the $\np$ interaction that binds the deuteron and that is used in \ref{subsec:d-c} 
and \ref{subsec:d-ni} (no absorption). 
The $\pBe$ is obtained from a direct fit to elastic data \cite{Lapouxsub}.  
We use the standard optical potential form
\begin{equation} \label{eq:3V(r)}
V(r) = - V_0 \; f(r, R_0, a_0) - i\,W_v f(r,R_v, a_v) ,
\end{equation}
where $f(r,R,a)$ is given by \Eq~\eqref{eq:f(r,R,a)}, plus the Coulomb interaction 
of a uniform charge sphere with radius $R_c = r_c A^{\frac{1}{3}}.$ 
A good fit to the $\pBe$ data at $E_\mathrm{Lab}/A = 39.1\MeV$ 
and up to $\theta_{\text{CM}}=70^{\circ}$ (see \Fig~\ref{be11elastic}) 
is obtained with the parameter set:
$V_0 = 51.2\MeV,$ $W_v = 19.5\MeV,$ $r_c = r_0 = r_v = 1.114\fm,$ 
$a_0 = 0.57\fm$ and $a_v = 0.50\fm,$ as used in \cite{Moro-06}. 
These parameters are slightly different from the ones proposed by 
Watson \etal~\cite{Watson-69}. 
Since the energy is sufficiently close to $\Bee\text{-}p$ scattering at $E_{\text{Lab}}/A = 38.4\MeV$ 
we expect the fit to be appropriate.

The Faddeev model space  contains partial waves $l \le 4$ in the $\np$ relative motion, 
$l \le 5$ in the $\nBe$ channel, and $l \le 22$  in the $\pBe$ channel. 
Again, this last channel is more demanding due to the presence of the Coulomb force. 
Total angular momentum up to $J = 20$ ($J = 40$ for breakup) is included.
The Coulomb potential is screened with a radius of $R=10\fm$
and smoothness $n=4$. An additional difficulty is the presence of the sharp $d$-wave
$\nBe$ resonance which is treated using the subtraction technique as in \Ref~\cite{Deltuva-06}.

CDCC calculations are performed using both the $\Bee$ breakup states (CDCC-BU), 
where the reaction mechanism is inelastic excitation of the projectile into its continuum, 
and the deuteron breakup states (CDCC-TR*), where the reaction mechanism involves transfer
to the continuum of the deuteron in the $\dpBe$ transfer channel \cite{Moro-06}. 
In the course of the calculations, it became apparent  that the model space used in 
\cite{Moro-06} was not enough to achieve full convergence of the CDCC-BU calculations. In this 
work, we have increased the number of partial waves for  the $\nBe$ relative motion,
as well as the number of multipoles for the coupling potentials,  
up to $\l_\mathrm{max}=8$, $Q_\mathrm{max}=8$, respectively.  
Even with this large number of partial waves, the results were not completely converged. Inclusion of 
higher partial waves led to numerical instabilities in the calculations, 
and hence the results presented here correspond to  $\l_\mathrm{max}=8$. Continuum bins
were calculated up to $E_{\mathrm{max}}=34\MeV$ for $l\le 6$ and  
$E_{\mathrm{max}}=32\MeV$ for $l=7,8$. 

For  CDCC-TR*, the model space is also augmented with respect to the 
calculations performed in
\cite{Moro-06}. The number of partial waves for the $pn$ relative motion is increased 
to $l \le 8$, and bins are considered up to  $E_{\mathrm{max}}=35\MeV$ and 
$R_{\mathrm{bin}}=60\fm$. We notice also that in the present CDCC-TR* calculations the 
n-$^{10}$Be interaction in the final channel (d+$^{10}$Be) is real, while in  \cite{Moro-06}
this interaction was complex.  
Multipoles $Q \le 4$ are included 
for the  CDCC coupling potentials. An extended non-locality range of 14~fm was used  for the 
transfer couplings.

For both the CDCC-BU and CDCC-TR* calculations the total angular momentum is  $J \le 35$, 
and the coupled equations are integrated  up to $R_{\mathrm{max}}= 60 \fm$. 


\section{Results} \label{sec:results}

Our three test cases are chosen to span a variety of situations. 
The $\dpC$ and  $\dpNi$ reactions at intermediate deuteron energies contain important effects that 
could not be well accounted for by simple prior form DWBA calculations \cite{Austern-87,Matsuoka-82}.
These reactions have been measured before and detailed breakup data 
are available for the former \cite{Matsuoka-82}.
Finally we include a reaction involving a loosely bound halo nucleus, the study of $\Beepp$, 
where three-body breakup has a decisive contribution and for which there have been 
several experimental studies (e.g. \cite{Lapouxsub}). 
Previous studies  \cite{Winfield-01}
have suggested  the relevance of the  interplay betweeen the 
breakup and the $(p,d)$ transfer channel. The possibility of including 
both  channels within the Faddeev formalism constitutes a further motivation 
of the present analysis. 

\subsection{$\bm\dpC$ at $\bm{E_d = 56\MeV}$} \label{results-c} 

\begin{figure}[t!]
\includegraphics[scale=0.35]{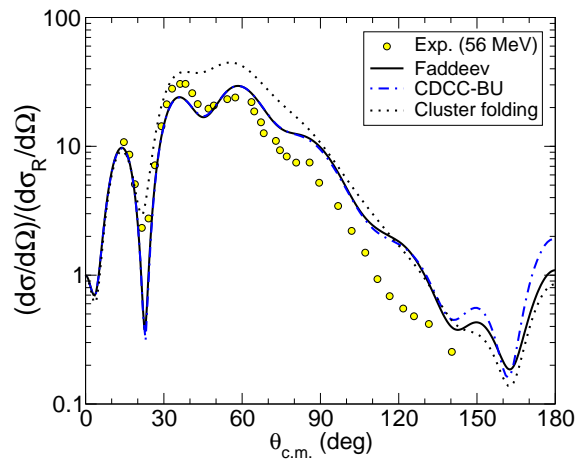}
\caption{\label{c12elastic} (Color online) Elastic cross section for deuterons on $\C$ at $E_d = 56\MeV$:
the solid line corresponds to exact three-body results and the dash-dotted line to CDCC. 
The result of a single channel cluster folding (dotted) is also shown.
The experimental data (diamonds) are from \Ref~\cite{Matsuoka-86}.}
\end{figure}

Our results for $\dpC$ elastic scattering at $E_d=56\MeV$ are presented in \Fig~\ref{c12elastic}, 
and compared to data around the same energy. The differential cross section $d\sigma/d\Omega$ is divided
by the corresponding Rutherford cross section $d\sigma_R/d\Omega$.
First we point out that the CDCC calculation (dash-dotted line) reproduces the exact three-body results 
(solid line) up to very large scattering angles. 
Secondly, there is agreement with the data at forward angles ($\theta_\text{CM} < 60^\circ$) 
but this agreement deteriorates for backward angles where mechanisms other than three-body breakup 
may start to play a role. 

In order to show the well known influence of the deuteron continuum on the elastic channel we have 
also included  in \Fig~\ref{c12elastic} the CDCC calculation without any coupling 
to the continuum (dotted line). This corresponds to a one-channel calculation 
with the deuteron-target potential given by the  single-folding 
expression:
$V_{00}(R)=\langle \phi_d | V_{pt}+V_{nt} | \phi_d \rangle$. It can be seen 
that this calculation largely overestimates the data in the angular 
region where the cross section is large, evidencing the importance of the deuteron 
breakup channel in the dynamics of the reaction. 

\begin{figure}[t!]
\includegraphics[scale=0.35]{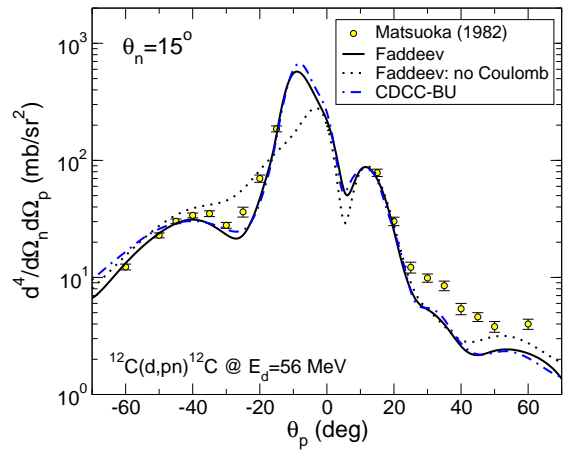}
\caption{\label{c12bu}  (Color online) 
Semi-inclusive differential cross section versus proton scattering angle
for the breakup of deuterons on $\C$ at $E_d = 56\MeV$: the solid line corresponds to
the exact 
three-body results and the dash-dotted line to CDCC. 
The dotted line corresponds to the three-body exact result in the absence of the Coulomb force.
The experimental data are from \Ref~\cite{Matsuoka-82}.}
\end{figure}

Next we consider the breakup observables. 
In \cite{Matsuoka-82}, measurements were taken by fixing the neutron detector at $\theta_n=15$ deg. 
We present both the proton angular distribution (after integration over energy) 
and the energy distributions for specific proton angles and compare to the data. 
The agreement between CDCC (dash-dotted line) and the exact three-body results (solid line)
is seen over all proton angles as shown in Fig.~\ref{c12bu}. 
Also shown in Fig.~\ref{c12bu} are the three-body results obtained without the Coulomb interaction 
(dotted line), where it becomes clear that Coulomb cannot be neglected for a wide angular range
at forward scattering angles. 
We note that the agreement of present CDCC calculations with the data has 
been  improved compared to the CDCC studies 
presented in \cite{Yahiro-84}, probably due to the larger model space included in our work.
\begin{figure*}[t!b]
\includegraphics[scale=0.5]{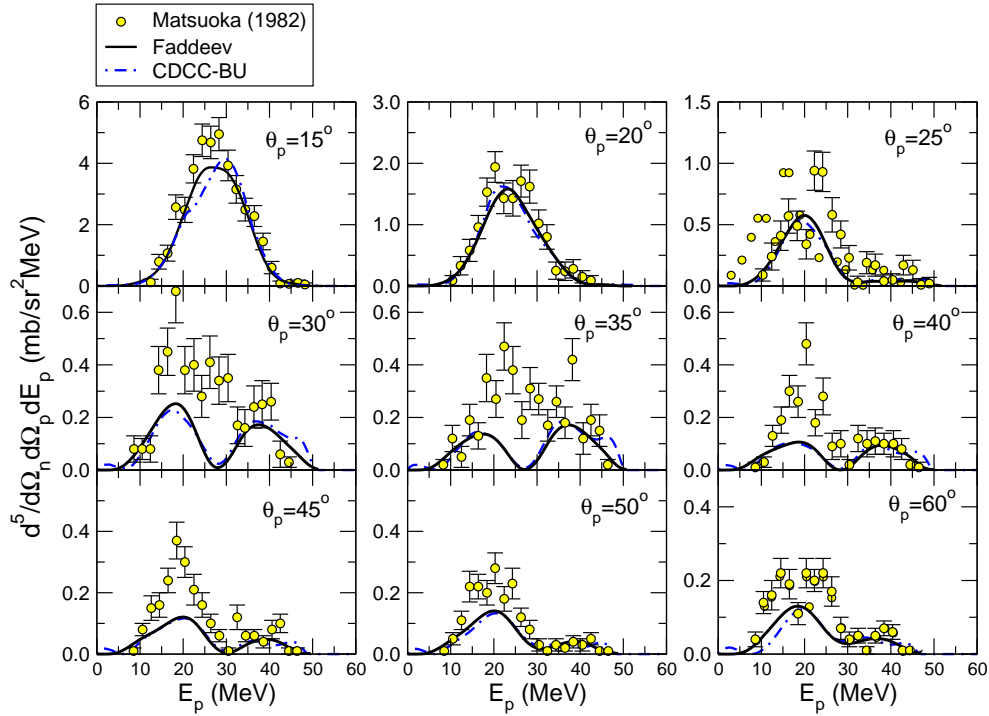}
\caption{\label{c12bu-d5p}  (Color online) Exclusive differential cross section versus proton energy for 
the breakup of deuterons on $\C$ at $E_d = 56\MeV,$ $\theta_n = 15^{\circ}$ and $\theta_p > 0$:
the solid line corresponds to exact three-body results and the dash-dotted line to CDCC.
The experimental data are from \Ref~\cite{Matsuoka-82}.}
\end{figure*}
\begin{figure*}[t!b]
\includegraphics[scale=0.5]{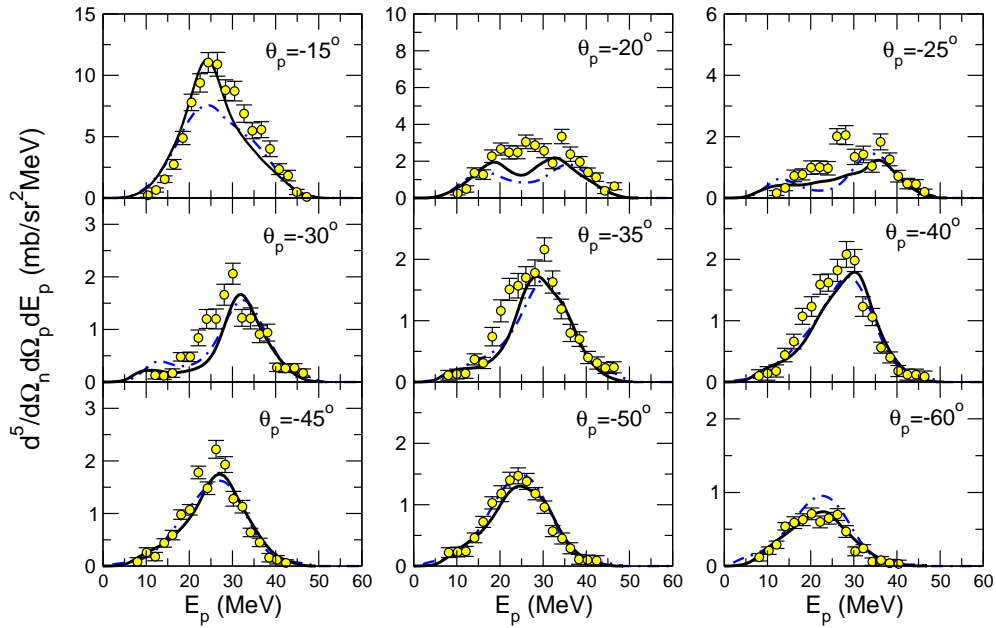}
\caption{\label{c12bu-d5m}  (Color online) Exclusive differential cross section versus proton energy for 
the breakup of deuterons on $\C$ at $E_d = 56\MeV,$ for $\theta_n = 15^{\circ}$ and $\theta_p < 0$:
the solid line corresponds to exact three-body results and the dash-dotted line to CDCC. 
The experimental data are from \Ref~\cite{Matsuoka-82}.}
\end{figure*}

In addition a note of caution needs to be added vis-a-vis the convergence of the three-body results 
with screening radius $R$ for $-30^\circ <\theta_p<10^\circ$ where we face the most demanding 
phase space constraints. 
Since this region corresponds to small  momentum transfer in $\pC$ subsystem,
the convergence with screening radius is slow and not uniform, forcing us to use 
a more sophisticated treatment for the breakup amplitude as described in the Appendix. 
For this reason the accuracy of our calculation is $10\%$ to $15\%$ for $-30^\circ <\theta_p<10^\circ$ 
and better than $5\%$ for all other values of $\theta_p$. 
The most sensitive region is the maximum of the cross section.

The proton energy distributions are shown in \Fig~\ref{c12bu-d5p} for protons being scattered
to the same side as the neutron ($\theta_p>0$), and in \Fig~\ref{c12bu-d5m} for protons coming out at 
opposite angles from the neutron ($\theta_p<0$). The overall agreement with the data is remarkable. 
Missing cross section is visible for the large positive scattering angles starting with 
$\theta_p=25^\circ$. As for the angular distribution, our CDCC results show an improvement compared to 
the analysis of \cite{Yahiro-84}, which we attribute to the inclusion of $l=4$ 
in the relative motion of $\np$ subsystem in our model space. 
Most important for this work is the realization that CDCC simulates the three-body effects 
contained in the solution of the exact three-body problem, even at this level of detailed observables. 
Again, for the reasons mentioned above, the three-body results shown in \Fig~\ref{c12bu-d5m} 
for $\theta_p = -15^{\circ}, -20^{\circ}$ and $-25^{\circ}$ may change slightly with the chosen 
screening radius while at all other angles we have fully converged results.

\subsection{$\bm\dpNi$ at $\bm{E_d = 80\MeV}$} \label{results-ni} 

\begin{figure}[t!]
 \includegraphics[scale=0.35]{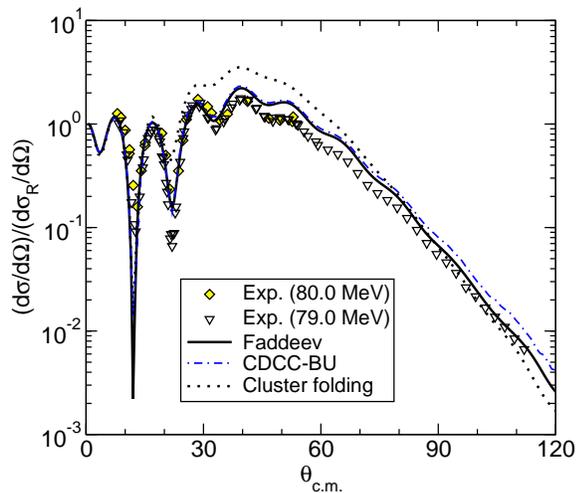}
\caption{\label{ni58elastic}  (Color online) 
Elastic cross section for deuterons on $\Ni$ at $E_d = 80\MeV:$ 
the solid line corresponds to exact three-body results and the dash-dotted line to CDCC results.
The experimental data at $80\MeV$ (diamonds) are from \Ref~\cite{Duhamel-71}, 
and those at $79\MeV$ (triangles) are from \Ref~\cite{Stephenson-83}. The 
dotted line is the CDCC calculation suppressing the coupling to the 
deuteron continuum (see text).}
\end{figure}

We have reproduced the elastic scattering results presented in \cite{Austern-87}, 
and compare in \Fig~\ref{ni58elastic} our theoretical predictions to two different sets 
of data around the same energy. 
As before, the solid line corresponds to the three-body results from the solution of the AGS equations, 
whereas the dash-dotted line is calculated using CDCC. 
Both calculations agree perfectly up to scattering angles $\theta_d=80^\circ$, 
providing  a good description of the data.
We also show the result with the cluster folding potential (dotted line) which demonstrates
the importance of deuteron breakup in the reaction mechanism, as was the case for
the elastic scattering of $d+^{12}$C.
The slight disagreement of the full calculations and the data could be due to the 
ambiguities of the 
optical potentials, or due to the influence of other 
channels not included explicitly in our calculations (such as transfer or 
target excitation). 

\subsection{$\bm\Beepp$  at $E_\mathrm{Lab}/A$=38.4 MeV} \label{results-be} 

\begin{figure}[t!]
\includegraphics[scale=0.35]{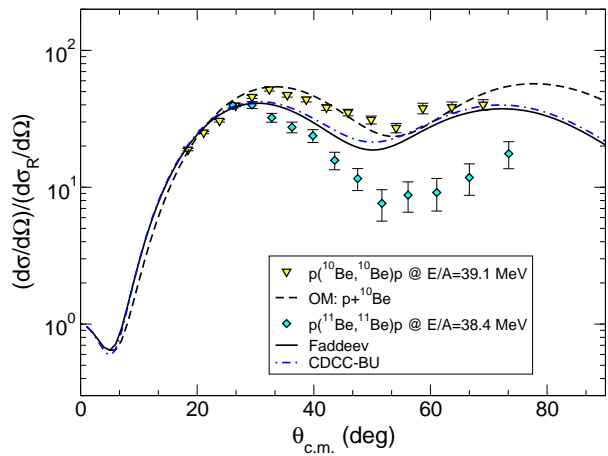}
\caption{\label{be11elastic}  (Color online) ${}^1{\mathrm H}(\Bee,\Bee)p$ elastic cross section at 
$E_{\mathrm{Lab}}/A = 38.4\MeV.$ 
The solid line corresponds to exact three-body results while the dash-dotted line to CDCC. 
The dashed line corresponds to an optical potential fit to the corresponding $\Be\text{-}p$ data 
of \Ref~\cite{Lapouxsub} shown by the triangles. 
The diamonds correspond to $\Bee\text{-}p$ elastic data of \Ref~\cite{Lapouxsub}.}
\end{figure}

Our last test case involves the breakup of the loosely bound $\Bee$ on a very light target, the proton.
Previous works have found difficulties in describing this process \cite{Summers-07} and,
furthermore, this reaction raised the red flag when comparing two different CDCC calculations 
which should produce the same results \cite{Moro-06}. 
We revisit the topic in the hope that the exact three-body calculations can help shed light 
on the issue. 

In \Fig~\ref{be11elastic} we show the results of our calculations for $\Beepp$ elastic scattering 
together with the corresponding data. 
For comparison we include $\pBe$ elastic data at $E_{\mathrm{Lab}}/A = 39.1\MeV$ 
and the corresponding theoretical fit (dashed curve) obtained with the $\pBe$ optical potential 
given in \Sect~\ref{subsec:pBe}.  
Two important features immediately arise: 
i) the agreement between the CDCC and the exact three-body results, and 
ii) the mismatch with the data.
In this work the first point is of more relevance than the second, demonstrating that
the CDCC takes well into account the three-body effects fully present in the AGS approach.
However, point ii) suggests 
that in this reaction, degrees of freedom beyond three-body breakup are being excited \cite{Summers-07}.

\begin{figure}[t!b]
\includegraphics[scale=0.4]{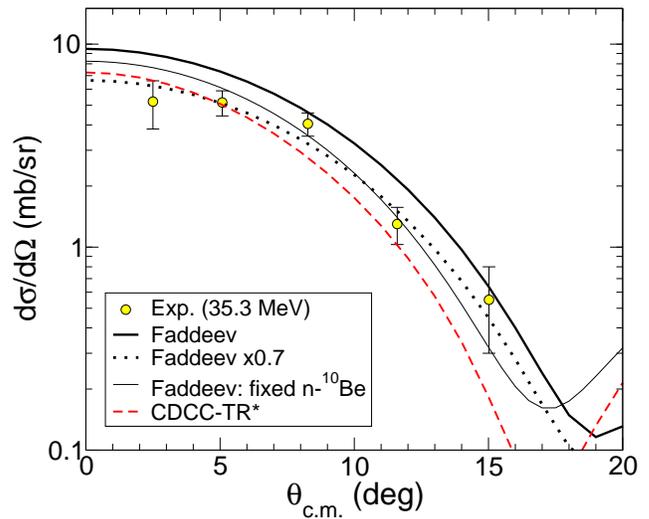}
\caption{\label{be11transfer}  (Color online) 
Transfer reaction ${}^1{\textrm H}(\Bee,\Be)d$ cross section at $E_{\mathrm{Lab}}/A = 38.4\MeV$.
The thick solid line corresponds to exact three-body result, while the  
dotted line corresponds to the same calculation multiplied by 0.7. The thin solid 
line is the exact calculation with a partial-wave independent $\nBe$ interaction. The latter is 
to be compared with the CDCC-TR* calculation (dashed line), as explained in the text. 
The experimental data are from \Ref~\cite{Winfield-01} at $E_p = 35.3\MeV.$}
\end{figure}

One main difference between this and the previous two examples is the explicit inclusion 
of the transfer channel in the three-body calculations. 
In other words, there is no absorption in $\np$ while the corresponding interactions 
in the previous two test cases $\nC$ and $\nNi$ included absorption.
In $\Beepp$, the neutron transfer channel is very important. 
We show in \Fig~\ref{be11transfer} the three-body predictions for the transfer 
$\Bee(p,d)\Be$ (thick solid line) together with the data for $E_p = 35.5\MeV$ \cite{Fortier-99}. 
The three-body calculation predicts the transfer cross section $\approx 20$\% above the data. 
If a simple proportionality of the transfer cross section to the square of the $\nBe$ $l=0$
single particle wavefunction were to be assumed, the three-body results would suggest
ground state spectroscopic factors consistent with previous works \cite{Winfield-01}. 

We also show in this figure the prediction of the CDCC-TR* (dashed line),
obtained with Eq.~(\ref{Tprior}). Due to the impossibility of including partial-wave dependent 
interactions in the evaluation of the coupling potentials, this calculation was performed assuming
$\nBe$ potential (\ref{eq:2V(r)}) with $V_0 = 51.639$ MeV in all partial-waves
that reproduces the ground state of $\Bee$.
 For a meaningful comparison with the exact result, 
we include also in this figure a Faddeev calculation performed with the same $\nBe$ interaction 
(thin solid line). Notice that in this case there is no bound excited state in $\Bee$. 
We see that the cross section for this CDCC-TR* calculation is about 15\% smaller than the AGS, 
and would hold a spectroscopy factor closer to unity. The  difference between
AGS and CDCC could be due to the fact that the CDCC wavefunction is not
a good reproduction of the exact three-body wavefunction in the surface region,
or that the choice of the optical potentials appearing in the remnant term
of Eq.~(\ref{Tprior}) is inadequate for this purpose,  which could be connected
to the poor description of the $\Bee$ elastic data. 
 
\begin{figure}[t!]
\includegraphics[scale=0.35]{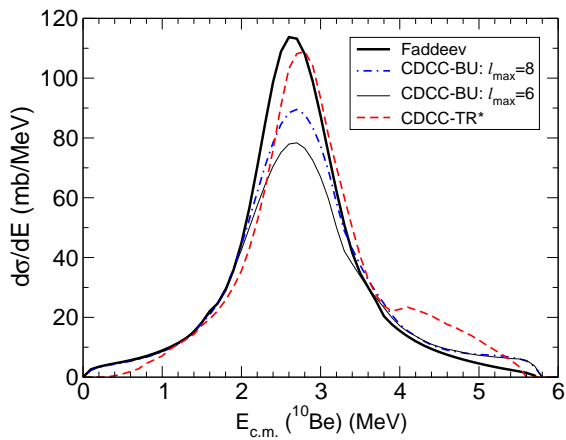}
\caption{\label{be11butrener}  (Color online) Semi-inclusive differential cross section for the reaction 
${}^1{\mathrm H}(\Bee,\Be)pn$, at $E_{\mathrm{Lab}}/A = 38.4\MeV,$ versus $\Be$ center of mass energy. 
The thick solid line corresponds to exact three-body results, the dashed line to CDCC-TR*,
the dash-dotted line and the thin solid line to CDCC-BU with $l_{max}=\mathrm{8}$ and  
$l_{max}=\mathrm{6}$, respectively.}
\end{figure}

\begin{figure}[t!]
\includegraphics[scale=0.35]{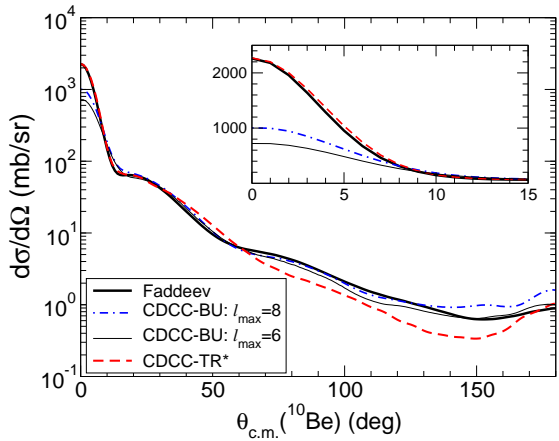}
\caption{\label{be11butrangl}  (Color online) Semi-inclusive differential cross section for the reaction 
${}^1{\mathrm H}(\Bee,\Be)pn$, at $E_{\mathrm{Lab}}/A = 38.4\MeV,$ versus $\Be$ angular distribution 
after energy integration. 
The solid line corresponds to exact three-body results, the dashed line to CDCC-TR* 
and the dash-dotted line and the thin solid line to CDCC-BU, with $l_{max}=\mathrm{8}$ 
and $l_{max}=\mathrm{6}$, respectively. }
\end{figure}

Finally in \Figs~\ref{be11butrener} and \ref{be11butrangl} we show the semi-inclusive differential 
cross section for the  breakup $\Bee+p \to \Be+p+n$ where $\Be$ is the detected particle. 
We present both, the energy distribution (\Fig~\ref{be11butrener}) and the angular 
distribution (\Fig~\ref{be11butrangl}). For the energy distribution, two CDCC-BU calculations 
are shown, one with $l \leq 8$ (dash-dotted line) and one with $l \leq 6$ (thin solid line) for 
the $\nBe$ motion. The significant 
difference between these two calculations suggests that the CDCC-BU calculation is not 
converged with respect to the number of $\nBe$ partial waves. The calculation 
with  $l \leq 8$  reproduces 
reasonably well the shape of the energy distribution predicted by the AGS calculation, 
but it underestimates this cross section 
at the peak by about 20\%. This underestimation could be due to the contribution of 
higher $\nBe$ partial waves or due to some breakdown of CDCC.
The angular distribution is also in good agreement with the exact result for the whole 
angular range, except at very small angles.

The CDCC-TR* calculation also reproduces reasonably well the energy distribution. 
For the higher $\Be$ energies, this calculation is however well above the 
AGS result. Moreover, some underestimation of the cross section is observed at the maximum 
of the distribution, as well as at low $\Be$ energies. We notice that these energies are 
associated with configurations in which the $pn$ system is in a very high excited state, 
and these states are difficult to include in the CDCC-TR* calculation. 
The peak observed in the energy distribution of \Fig~\ref{be11butrener} corresponds 
to $\np$ quasi-free 
scattering and it is natural that it is best reproduced by CDCC-TR*.
Note that, as for the CDCC-TR* predictions for the transfer to the ground state, 
in the CDCC-TR* breakup, the fixed $\nBe$ partial-wave independent interaction 
was used to generate the deuteron continuum. However in this case the results
are not very sensitive to this potential choice. 

The CDCC-TR* angular distribution 
(dashed line in Fig.~\ref{be11butrangl}) reproduces well the Faddeev calculation at small 
angles, but underestimates the cross section for angles beyond 50$^\circ$. We notice again 
that small angles are mainly associated with the $np$ quasi-free scattering region
which is better described in a $pn$ basis, as it is done
in the TR* approach.  Conversely, these configurations are difficult to describe in the 
$\nBe$ basis which explains the low convergence rate of the 
BU calculation at small $\Be$ scattering angles. 

We finish this section by noting that, even if the  CDCC wavefunction is not appropriate 
to describe breakup in all regions of phase-space, this wavefunction can be accurate in a 
limited domain. For example, in the CDCC-BU calculations presented in this section, the CDCC solution 
can be a good approximation to the exact three-body wavefunction in the region of space that corresponds 
to small neutron-$\Be$ separations.


\section{Conclusions} \label{sec:conclusions} 

A comparative study of reaction observables calculated within the three-body AGS framework
and the approximate CDCC equations is presented. 
The AGS results shown here involve heavier nuclei 
where optical potentials together with the full treatment of the Coulomb interaction are used to describe 
direct nuclear reactions driven by deuterons and halo nuclei. 
We perform calculations for the scattering of deuterons on $\C$ at $E_d=56\MeV$ and $\Ni$ at $E_d=80\MeV,$ 
as well as $\Bee$ on protons at $E_{\text{Lab}}= 38.4\MeV/A,$ and calculate elastic, 
breakup and transfer observables.

The results indicate that, for reactions involving the elastic scattering and breakup of deuterons on 
carbon and Ni targets, CDCC is in agreement with the full three-body results.
Our calculations also reveal that Coulomb effects in the breakup of deuteron by $\C$ 
are not negligible for proton forward-angle breakup. 
Indeed, the method of screening and renormalization used to treat the Coulomb interaction 
in the AGS equations is stretched  to its limit of applicability in these regions of 
phase space characterized by small momentum transfer in the $\pC$ subsystem, 
a situation never encountered before in $\pd$ 
\cite{Deltuva-05-05-05} or $d\text{-}\alpha$ breakup \cite{Deltuva-06}.

For the $^{11}$Be-proton test case, the picture is more complicated.
For elastic scattering the two methods are in good agreement, however still fall short
to describe the data. For the transfer cross section, CDCC constructed on the deuteron 
continuum underestimates the cross section compared to the solution of the AGS equations.
And finally for the breakup observables, we only find good agreement between CDCC and Faddeev 
in certain regions of phase space, depending strongly on the choice of the basis used
for the CDCC expansion. Specifically, in the energy regime  
dominated by $\pn$ quasi free scattering (forward $^{10}$Be angle) the representation 
based on the $pn$ system (CDCC-TR*) accounts well for the full three-body effects, 
whereas for large angles of the detected fragment,
corresponding to small excitations of the $n-^{10}$Be system, the basis constructed
from the $^{11}$Be continuum (CDCC-BU) is more appropriate. A word of caution is required
for the choice  of the CDCC basis to be used for given kinematical regimes. It is
also important to note that the rate of convergence of the CDCC observables 
is very slow, particularly for CDCC-BU.
This may in part explain the disagreement found at some angles and energies of the detected
fragments.

The CDCC equations attempt to produce a wavefunction that describes all the
three-body effects, from small internal projectile distances to very large ones.
In fact, all CDCC-BU observables are obtained here from  the asymptotics of the three-body
wavefunction. Since this is computationally very demanding, an alternative has been suggested that
consists in using the wavefunction only in the range of the interactions, that is, inserted
into the post form transition amplitude \cite{johnson-ria}. Further work on this topic
is needed to explore such possibility.

\begin{acknowledgments}
A.D. is supported by the Funda\c c\~ao para a Ci\^encia e a Tecnologia (FCT) grant SFRH/BPD/14801/2003,
E.C. and A.C.F. in part by the FCT grant POCTI/ISFL/2/275, A.M.M. by 
Junta de Andaluc\'{\i}a and by the Ministerio de Eduaci\'on 
y Ciencia under project FPA2006-1387-C,
and F.M.N. by the National Science Foundation through grant PHY-0555893. We 
are grateful to Jeff Tostevin, for providing us the code to calculate 
breakup observables in CDCC calculations. We thank Ron Johnson for useful discussions
related to CDCC.
\end{acknowledgments}

\begin{appendix}
\section{}

The kinematical situations characterized by small momentum transfer $\Delta k$ in the subsystem
of charged particles are sensitive to the screened Coulomb potential at large distances.
In elastic scattering $\Delta k$ may even vanish, but the problem is resolved by
separating the long-range part of the amplitude \eqref{eq:Tcm} and explicitly
performing the $R \to \infty$ limit.
In breakup the transition operator is a Coulomb-distorted short-range operator, $\Delta k$ is always 
nonzero, and the $R \to \infty$ limit can be reached at finite $R$ with sufficient accuracy.
Nevertheless, the decomposition of the breakup operator \eqref{eq:U0a}
into two parts with different range properties 
\begin{gather}
U^{(R)}_{0 \alpha}(Z) = B^{(R)}_{0 \alpha}(Z) + [U^{(R)}_{0 \alpha}(Z)-B^{(R)}_{0 \alpha}(Z)]
\end{gather}
may be useful in practical calculations. It was shown in \Refs~\cite{Deltuva-05-05-05,Alt-78-80,alt:94a}
that the $R \to \infty$ limit in \Eq\eqref{eq:UC0} exists for both parts separately and that the
longer-range part of the breakup amplitude is given by
\begin{gather} \label{eq:B0a}
B^{(R)}_{0 \alpha}(Z) = [1 + T_{\rho R}(Z) G_{0}(Z)] v_\alpha
            [1 + G^{(R)}_{\alpha}(Z) T^{\cm}_{\alpha R}(Z)]
\end{gather}
where $\rho$ is the neutral particle and
 $T_{\rho R}(Z) = w_{\rho R} + w_{\rho R} G_0(Z) T_{\rho R}(Z)$ is the two charged particle 
screened Coulomb transition matrix. This part of the amplitude, called pure Coulomb breakup term,
requires larger screening
radius for convergence if $\Delta k$ is small, but is simpler to calculate than the full 
$U^{(R)}_{0 \alpha}(Z)$. In contrast, the convergence with $R$ is faster for the
shorter-range part $[U^{(R)}_{0 \alpha}(Z)-B^{(R)}_{0 \alpha}(Z)]$,
even when $\Delta k$ is small.
Therefore in kinematical configurations of $\dpC$ breakup with small $\Delta k$ it is sufficient to
calculate $[U^{(R)}_{0 \alpha}(Z)-B^{(R)}_{0 \alpha}(Z)]$ with standard parameters described in 
Sec.~\ref{subsec:d-c}, but the remaining term $B^{(R)}_{0 \alpha}(Z)$ needs considerably larger $R$. 
In the latter case we use the form
\begin{gather} \label{eq:B0aw}
\begin{split}
B^{(R)}_{0 \alpha}(Z) = {}& w_{\rho R} G_0(Z) T_{\rho R}(Z) - 
W^{\cm}_{\alpha R} G^{(R)}_{\alpha}(Z) T^{\cm}_{\alpha R}(Z) \\ 
& + T_{\rho R}(Z) G_{0}(Z) v_\alpha G^{(R)}_{\alpha}(Z) T^{\cm}_{\alpha R}(Z) \\
& + w_{\rho R} - W^{\cm}_{\alpha R}
\end{split}
\end{gather}
that is equivalent to \eqref{eq:B0a} on-shell.
The partial-wave convergence is slowest for the last term $w_{\rho R} - W^{\cm}_{\alpha R}$
which we therefore calculate without partial-wave expansion. 
Reasonably converged $\dpC$ breakup results in the present paper are obtained
with the screening radius up to $60\fm$ for the pure Coulomb breakup term.
Partial waves with $\pC$ orbital angular momentum $l \leq 38$ and total angular momentum
$J \leq 100$ are included in the calculation of $B^{(R)}_{0 \alpha}(Z)$.

\end{appendix}



\end{document}